# Indoor Navigation on Google Maps and Indoor Localization Using RSS Fingerprinting

[1]Sagar V. Ramani, [2]Yagnik N. Tank.

*[1,2]Lecturer(Computer Engg .Department ), Government Polytechnic , Porbandar , Gujarat, India*

*Abstract*— **Contrasting to advances in street/outdoor navigation, wall mounted maps and signs continue to be the primary reference indoor navigation in hospitals, malls, museums, etc. The proliferation of mobile devices and the growing demand for location aware systems that filter information based on currently device location have led to an increase in research and product development in this field. An attempt has been made to provide solution for indoor navigation on Google maps and to provide location of a user in a building using Wi-Fi signal strength on android Smartphone.**

## I. INTRODUCTION

Built information is a central factor in our life and a big portion of human life is spent inside buildings. There are many types of indoor and outdoor methodologies are available. A majority of these solutions are based on Global positioning system (GPS) and instant video and image processing. An indoor navigation can be assisted by using the building maps and plans documented in computer-aided design applications (CAD).

With the trend towards ubiquitous computing, context awareness is becoming a key factor in multiple applications, with location information playing a fundamental role. Most recent studies to monitor locationinformation obtained through GPS is only reliable in certain outdoor environments with direct visibility to the satellites.

Using CAD information, indoor floor plans can be drawn containing rooms, hallways, stairs, etc. Google maps and the API's are easily available to be molded on mobile devices, Floor plans can be overlaid on Google maps and can be used for navigation. For Localization, Wi-Fi signal strength can be used to detect the location of a user inside building.

Within the emerging market of smart phones, Android Operating System is well suited for modern devices and offers an excellent option for localization due to its support of diverse sensors and hardware which can be leveraged for localization beyond GPS (e.g. Wi-Fi, cellular communications, accelerometer), and its higher efficiency in comparison with other platforms such as J2ME [1].

In section 2, related work has been summarized. Section 3 describes proposed model and conclusion is given in section 4.

## II. RELATED WORK

The research on indoor navigation systems has been driven greatly by navigation utilizing images and embedded cameras on mobile devices has been a subject of studies. These include a system which delivers photorealistic panoramic images using 3D models to mobile phones, and a system which determines the location from images taken from point markers [2, 3] , as well as new types of concepts, such as Rotating Compass, a service, which guides users through public displays in-collaboration with mobile phone [4], have emerged. Systems and concepts helping people to navigate in extreme situations, e.g., how fire-fighters could benefit from indoor navigation [5], are being developed. In addition new indoor navigation UI designs for the physically impaired have been researched, such as in [6]. To the best of our knowledge, our indoor navigation application on Google maps is the first one on smart phones.

Radio Signal Strength Indications (RSSI) can be translated into distances from beacon points by means of theoretical or empirical radio propagation models. The following expression accounts for a general radio propagation model delivering the received power Pr.

$$Pr = Pt \ \{\lambda / 4\Pi d \ \}nGt \ Gr$$

Where Pt represents the transmitted power, the wavelength of the radio signal, Gt and Gr the gains of the transmitter and receiver antennas respectively, d the distance separating them, and n is the path loss coefficient, typically ranging from 2 to 6 depending on the environment. The two main approaches for the estimation of location making use of RSSI values are: 1) "fingerprinting", where a pre-recorded radio map of the area of interest is leveraged to infer locations through best matching, and 2) "propagation based", in which RSSI values are used to calculate distances through the computation of the path loss. "Propagation based" techniques can face errors of up to 50% [7] due to multipath, non line-of-sight conditions, interferences and other shadowing effects, rendering this technique unreliable and inaccurate, especially for indoor environments, where multipath is very important.

The fingerprinting technique has already shown promising results [8]. The fingerprinting process consists of two phases:

### A. *Phase-1*

Training Phase or offline phase, in which a radio map of the area in study is built. RSSI values from different beacons are recorded at different locations; the separation between these chosen locations will depend on the area in study, and for instance, for indoor environments this separation can be of around a meter [9]. Each measurement consists of several readings, one for each radio source in range.

B. *Phase-2*

Online phase, in which the mobile terminal infers its location through best matching between the radio signals being received and those previously recorded in the radio map.
Regarding Wi-Fi technology, several research groups have already tried to leverage RSSI fingerprinting for localization:
Radar: Represents the first fingerprinting system achieving the localization of portable devices, with accuracies of 2 to 3 meters.

- Horus: based on the Radar system, it manages a performance improvement making use of probabilistic analysis.
- Compass: applies probabilistic methods and leverages object orientation to improve precision, claiming errors below 1.65 meters.
- Ekahau: commercial solution using 802.11 b/g networks, achieving precisions from 1 to 3 meters in normal conditions.

Nevertheless, all the existing approaches use dedicated and complex hardware, making them unfeasible for direct implementation in smart phones.
The RSS fingerprinting technique for localization can be utilized with other radiofrequency technologies including:

- Bluetooth, which despite the extra infrastructure requirements in comparison with Wi-Fi, it can achieve accuracies in the range of 1.2 meters.

- Conventional radio can also be used for localization. However, the requirement of dedicated hardware and the fact that devices can be located only down to a suburb, represent important drawbacks.

- Digital TV signals have also proved to be suitable for localization, but subject to dedicated hardware requirements and low resolutions.

### III. PROPOSED MODEL

A. *Indoor Navigation on Google maps:*

We have studied the possibility of providing indoor navigation on Google maps. The two floors of F-block, PESIT, Bangalore were considered for providing navigation. The floor plans were drawn on AutoDesk Online IDE using the blueprint of building. The aim was to exactly overlay the floor plans of building on the Google maps so that Geocodes (Latitude, longitude) information can be accurately collected for the rooms, Corridors, Stairs, etc. The building elements (rooms, corridors, stairs, etc.) are considered as nodes (source or destination). The edges are link between different nodes. Dijkstra's algorithm was considered for providing shortest path between nodes. The Google map and android application can be integrated using Google map API key. The Geocodes were collected using Google maps, Geocodes represent the nodes on the floor plan. The Dijkstra's algorithm sends the shortest path, the edge is drawn between nodes using the geocodes.

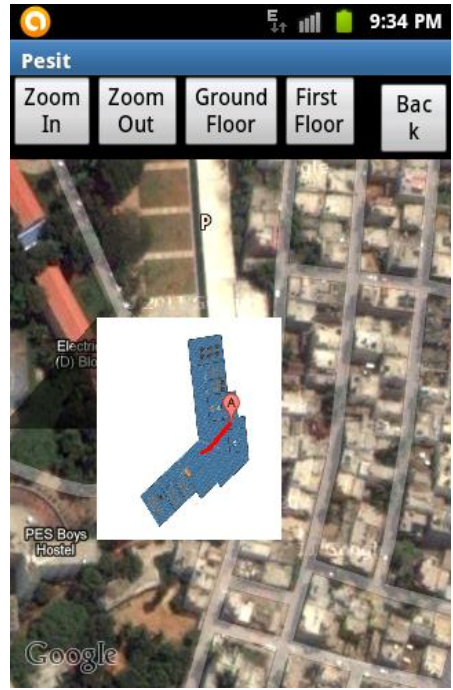

Fig 1: Floor Plan on a Google map

B. *Indoor Localization using RSS fingerprinting*

RSS information from Wi-Fi beacons deployed within buildings allows us to obtain a radio map of different locations (Technique called fingerprinting), and we will estimate locations through the comparison of the current RSS measurements with those stored in the radio map. Different attempts to obtain RSS-based indoor localization without fingerprinting show an important loss of accuracy[10].Also, many fingerprinting-based localization systems make use of dedicated hardware for the collection of data in the training phase, while in the measurement phase, the actual mobile device used for localization is different, resulting in an error called "signal reception bias" [11], due to the differences in antennas characteristics and measurement acquisitions schemes between different equipment. We have carried out tests to measure different radiofrequency signal strengths within the PESIT F-block. Wi-Fi technology offers the most reliable approach for indoor localization in our building, because of the important deployed infrastructure of Wi-Fi Access Points providing coverage in the whole building. For the measurement of the signals and practical implementation of our localization application, we have used smart phones running on Android.In our experimental setup, each Wi-Fi Access Point has 5 radios (each represented by a MAC address).

International Journal of Engineering Trends and Technology (IJETT) – Volume 11 Number 4 - May 2014

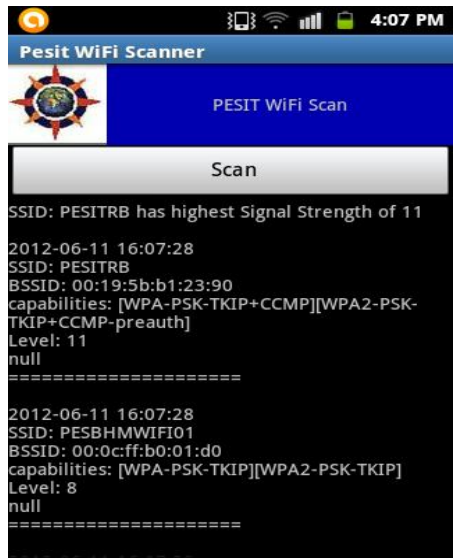

Fig 2: RSS logger Application

For example, 00:19:5b:b1:23:90, 00:23:04:89:24:08, 00:23:04:89:1f:98, 00:23:04:89:67:e7 and 00:23:04:89:22:00 are 5 radios belonging to the same Access Point(PESITRB). RSS values (in dBm) from the same Access Point can show important standard deviations in between consecutive scans (within the same radio) and also in between different radios within the same Access Point. Consequently, averaging of values both within the same Access Point and over time provides much more stable values that can successfully be used as a fingerprintcomponent of each particular location. We call this approach "Nearest Neighbour in signal space and Access Point averages".It must also be noticed that there are radios which only show up at some particular locations, but with a very low RSS value. For instance, 00:0c: ff: b0:01:d0 could only be noticed in a small area in our building with 3 dBm. Nevertheless, this information with so low RSS values can be considered as noise, and therefore it is not reliable to use radios with RSS below 15 dBm as a disambiguation factor. The useful information to disambiguate locations can be found at RSS values above 15 dBm, which in our case corresponds to approximately 3 Access Points with stable enough values to allow us to correctly distinguish between the different locations.

TABLE I
FEW RADIO MAPS DATA OF AUDITORIUM

| SR NO. | SSID | BSSID | RSSI( dBm ) |
|---|---|---|---|
| 1 | PESITRB | 24:08 | -62 |
| 2 | PESITRB | 23:90 | -80 |
| 3 | PESITRB | 1f:98 | -53 |
| 4 | CISCO_LAB | 67:e7 | -85 |

TABLE II
FEW RADIO MAPS DATA OF ISE DEPT. OFFICE

| SR NO. | SSID | BSSID | RSSI( dBm ) |
|---|---|---|---|
| 1 | PESBHMWIFI01 | 27:78 | -85 |
| 2 | PESITRB | 23:90 | -86 |
| 3 | PESBHMWIFI01 | 1a:90 | -68 |

## IV. CONCLUSION

We have developed an approach to provide indoor navigation on Google maps. This application can be used in any huge buildings where indoor navigation is necessity. An indoor localization application to be implemented in smart phones, leveraging their sensing capabilities in order to deliver up to 1.5 meters accuracy without the requirements for complex hardware that existing solutions need. This application delivering up to 1.5 meters resolution making use of only the hardware embedded within the phone and integrating both online and offline phases of RSSI fingerprinting within the same device.